\documentclass[1p,12pt]{elsarticle}
\usepackage{graphicx}
\graphicspath{{./}}
\usepackage{amsmath}
\usepackage{amssymb}
\usepackage{lineno}
\usepackage{textcomp}

\linenumbers
\journal{Nucl. Instr. and Meth. A}

\begin{document}

\begin{frontmatter}

\title{Simultaneous reconstruction of scintillation light and ionization charge produced by 511 keV 
photons in liquid xenon : potential application to PET}
\author[a]{P.~Amaudruz}
\ead{amaudruz@triumf.ca}
\author[b]{D.~Bryman\corref{cor1}}
\ead{bryman@phas.ubc.ca}
\author[a]{L.~Kurchaninov}
\ead{kurchan@triumf.ca}
\author[b]{P.~Lu}
\ead{philipfl@phas.ubc.ca}
\author[a]{C.~Marshall}
\ead{cammarsh@triumf.ca}
\author[c]{J.~P.~Martin}
\ead{jpmartin@lps.umontreal.ca}
\author[a]{A.~Muennich}
\ead{muennich@triumf.ca}
\author[a]{F.~Retiere}
\ead{fretiere@triumf.ca}
\author[a]{A.~Sher}
\ead{sher@triumf.ca}
\cortext[cor1]{corresponding author,  Phone 001-604-222-7338, Fax 001-604-222-1074}

\address[a]{TRIUMF, 4004 Wesbrook Mall, Vancouver, BC, V6T 2A3}
\address[b]{Department of Physics and Astronomy, University of British Columbia, 6224 Agricultural Road, 
Vancouver, BC, Canada V6T 1Z1}
\address[c]{University of Montreal, CP 6128 Succursale Centre-Ville, Montreal, Quebec, H3C 3J7 Canada}

\begin{abstract}

In order to assess the performance of liquid xenon detectors for use in positron 
emission tomography we studied the scintillation light and 
ionization charge produced by 511~keV photons in a small prototype detector. Scintillation light 
was detected with large area 
avalanche photodiodes while ionization electrons were collected on an anode instrumented with low 
noise electronics 
after drifting up to 3~cm. Operational conditions were studied as a function of 
the electric field. Energy resolutions of $<10$\% (FWHM) were achieved by 
combining the scintillation light and ionization charge signals. The relationship between scintillation 
light and ionization signals was investigated. An analysis of the sources of fluctuations was performed 
in order to optimize future detector designs.

\end{abstract}


\end{frontmatter}

\section{Introduction}
Positron Emission Tomography (PET) is a functional imaging technique of growing importance in medical 
diagnostics.  Its powers lie in the ability to reveal biologically significant processes that can be
used, for example, in cancer screening and  in studying neurodegenerative diseases.  
Conventional PET detectors employ scintillating inorganic crystals~\cite{PET} as the gamma ray  
detection media.  While crystal-based PET systems perform adequately for many applications there 
is motivation for seeking improvements of  resolutions in energy, position, and time response to improve image quality
and increasing overall sensitivity.  
Liquid xenon (LXe) is another gamma ray detector technology~\cite{book} applicable to  high resolution PET which may 
result in improved performance and reduced noise in 
images due to superior energy resolution, true 3-dimensional position reconstruction, and the capability for 
determining the Compton scattering sequence~\cite{comptonrec1, comptonrec2, comptonrec3}.
Energy resolution of 7\% (FWHM) has been reported in small LXe detector tests by combining scintillation 
light and ionization charge measurements~\cite{columbia}.
Measuring charge in a drift chamber has been 
shown to provide 3-D sub-millimeter spatial  resolution~\cite{lxegrit, solovov} 
because electron diffusion is very small~\cite{ediff}. In addition, sub-ns timing resolution has been 
achieved by measuring the scintillation light~\cite{timing}.  Liquid xenon is also inexpensive 
compared to crystal detectors commonly used for PET. Liquid xenon PET systems have the potential to 
reduce detector contributions to PET to the level of intrinsic limitations due to positron 
range and non-colinearity of the emitted photons. 

This paper deals with the energy resolution obtained from light and charge signals observed in a small 
LXe prototype detector as well as an investigation of  the components influencing it and the sources 
of uncertainty which may inform the design of future detectors for PET.

\section{Micro-PET Detector Design}
\begin{figure}[!h]
\begin{center}
\includegraphics[width=0.6\linewidth]{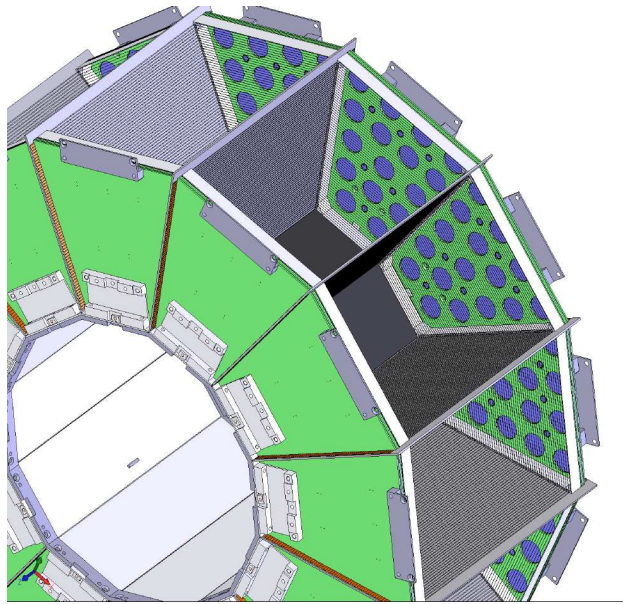}
\caption{The LXe PET ring concept. Scintillation light and charge are measured in each of the 12 
modules consisting of a LXe time projection chamber viewed by avalanche photodiodes.}
\label{fig_PET}
\end{center}
\end{figure}

We have developed a concept for a micro-PET detector shown in Fig.~\ref{fig_PET} that takes 
advantage of all the high resolution capabilities of LXe gamma ray detectors. Scintillation light is 
measured by arrays of large area avalanche photodiodes (LAAPD), which have 
been found to work well in LXe~\cite{LAAPD}. Charge measurement is achieved by using 
a time projection chamber (TPC), an approach successfully demonstrated  
in~\cite{lxegrit}.
Photons entering the LXe produce prompt scintillation light  and ionization  which drifts under an 
electric field applied between the cathode and the anode of the TPC.  The anode module 
(not shown in Fig.~\ref{fig_PET}) consists of a shielding grid followed by an array of wires preceding 
the anode which is  segmented 
into strips perpendicular to the wires. The electron signal induced on the wires and collected by the
strips provides a two dimensional (x-y) position measurement of the charge. The third coordinate (z) 
is obtained by measuring the electron drift time i.e. the difference between the time of the light flash
and the electron arrival time on the anode. Since every interaction is 
precisely recorded, Compton scattering can be reconstructed giving information on the 
direction of each incoming photon providing the possibility to suppress accidental coincidences and 
scattering prior to reaching the detector.

The expectations for performance under operating conditions for PET include  sub-millimeter 3-D position 
resolution from charge, 
timing resolution of  $<1$~ns from scintillation light, energy resolution $<10$\% (FWHM) combining 
light and charge signals, and the ability to reconstruct Compton scattering. 
Spatial location of events obtained from the prompt  distributed light signals will be used  
to reduce the ambiguities of  associating the scintillation light and charge at high levels of activity. 
A simulation of the imaging performance of this system will be presented in a future publication~\cite{simpaper}.

\section{Small Chamber Prototype}
\subsection{Test Setup}
\begin{figure}[!h]
\begin{center}
\includegraphics[width=\linewidth]{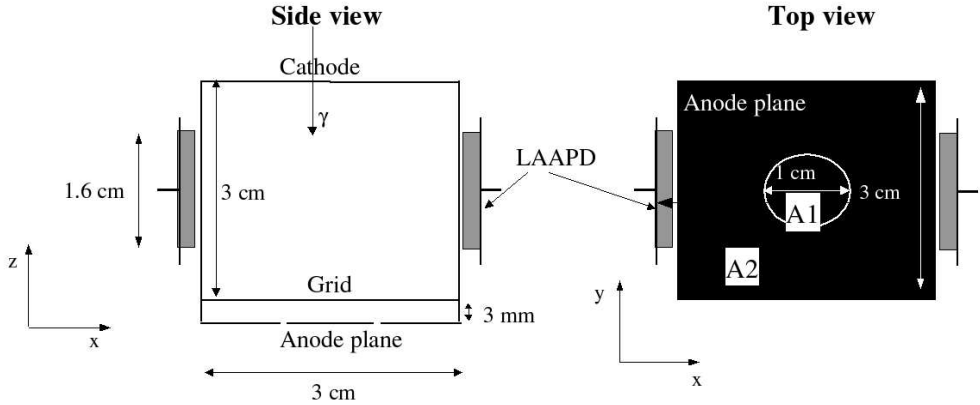}
\caption{Schematic views of the small test chamber. The side view illustrates the drift direction 
between the cathode and anode, viewed by two LAAPDs immersed in the LXe. The top view shows the 
segmentation of the anode.}
\label{fig_chamber}
\end{center}
\end{figure}

As an initial step in studying LXe detectors for PET,  we constructed  a small test chamber (27~cm$^3$)
for simultaneous  measurements of  light and charge.
The test chamber is shown schematically in Fig.~\ref{fig_chamber}.

Scintillation light was detected by two 1.6~cm diameter windowless LAAPDs (Advanced Photonics Inc.~\cite{API}). 
The LAAPDs were located 
at the center of the drift region as  shown in Fig.~\ref{fig_chamber}, 1.5~cm above the grid wires and 1.5~cm below the cathode. 
Charge was collected on a central 1~cm diameter electrode (A1) or on an outer electrode (A2).
An electric drift field was applied between the cathode and a shielding grid separated by 3~cm.
The electric field was formed by a field cage consisting of 9 wires with a spacing of 3 mm
strung along the four walls of the chamber. The voltage was distibuted by 100M$\Omega$ resistors.
The APDs were outside the field cage and the distance between the field cage wires and the APDs was 2mm.
The shielding grid consisted of 0.1~mm dia. wires spaced by 3~mm located 3~mm from 
the anode charge collection plane. 
The electric field between 
the grid wires and the anode was set  higher than the drift field ensuring that all the 
electrons pass through the grid~\cite{grid}. In order to study the influence of the drift field on
quantities like charge and light production and the energy resolution several settings were used.  
With the grid at ground potential measurements were made with the negative cathode voltage set to 1, 3, 6, 
and 8~kV, with the respective anode voltages set to 300, 600, 1200 and 1200~V. Photons of
511~keV emitted after annihilation of positrons from a $^{22}$Na source with an activity of $9.61\cdot 10^5$ Bq and situated in a collimator with an opening angle of 2$^\circ$ positioned 30~cm away from the cathode
entered the test chamber (along the z axis) through the cathode plane. The trigger was generated by selecting signals in coincidence 
of both APDs and an external NaI detector placed at a distance of 50 cm from the source observing the full energy of the other 511~keV photon 
from the positron annihilation. The probability to detect more than one event in the chamber at the same time was less than 3\%.
The detector was operated at 15 psia and at temperatures between 168 and 169 K. 
Before inserting the liquid in the vessel holding the detector a bake-out in vaccuum at 7.6$\cdot 10^{-6}$T was performed for 6  days at 60$^\circ$C
to clean the components. 
The purification of the xenon was done in the gas phase using
two stages both with equipment from From NuPure Corporation~\cite{nupur}: 
first, the heated getter (NuPure Omni 600) was used to remove H$_2$0, O$_2$, CO, H$_2$, and N$_2$ to sub-ppb levels
followed by a room temperature getter (Eliminator 600 cg) to remove H$_2$0, O$_2$, CO, H$_2$, 
and hydrocarbons to $<$ 0.5 ppb. The lifetime of drifting ionization electrons was used to indicate successful operation of the purification as discussed below.

\subsection{Readout Electronics}
The two anodes segments  and the grid wires (ganged together) were connected to charge-sensitive 
amplifiers followed by a 1~\textmu s time constant RC-CR shaper. 
The amplifier was calibrated using a narrow pulse input charge with a precision of 5\%. The amplifier outputs were 
fanned out into three branches:
\begin{enumerate}
\item  A constant fraction discriminator followed by a time-to-digital converter (TDC) CAEN model V1190B;
\item  A charge sensitive analog-to-digital converter: 12 bit QDC CAEN model V792 
  with gate adjusted to the drift time and pulse shape; and
\item  A 20~MHz sampling waveform digitizer VF48~\cite{vf48}.
\end{enumerate} 
To get absolute charge values, the digitized waveform measured with the VF48 was used for the analysis 
presented in sections \ref{sec_Q} and \ref{sec_L}.
Because the QDC had a better signal to noise ratio it was used to determine the energy 
resolution in section \ref{sec_E} which did not require absolute charge calibration.
The other reason the QDC was not used for absolute values was due to the very short pulse used for 
calibration. The longer chamber signal would not have been fully integrated within the 
window set.

The observed range of noise of the amplifier  was 700-1100 electrons due to varying external sources of induced noise. 
To reach the optimal position resolution, a signal to noise ratio larger than 5 was desirable 
requiring the electronics noise to be kept below 1000 electrons equivalent noise 
charge (ENC). A typical signal was expected to be  at least 10 000~e-, as long as the electron 
attachment during the drift was small.

The LAAPD 
voltages were set so that their gains were  500 and each was connected to a current-sensitive 
preamplifier with a pulse width of 50~ns and  
$10^4$ electrons ENC. The amplifier signal was split into 3 branches: 
\begin{enumerate}
\item Discriminator and TDC;
\item QDC CAEN model V792 with a gate of 100~ns;  and
\item  1~GHz waveform digitizer CAEN model V1729. 
\end{enumerate} 

Solid angle calculations showed that 12\% of the scintillation
photons reached the LAAPDs when the gamma interaction took  place in the center of the chamber. 

\section{Charge Collection}\label{sec_Q}
The grid wires shielded the anodes from the current induced by the drifting electrons.  Once the 
ionization electrons passed the grid wires the signal on the anode started to build
up with a pulse shape that was largely independent of the z position of the primary interaction 
although electrostatic calculations showed that the current pulse shape depended on the x-y
distance of the electron cloud from the individual grid wires. Furthermore, depending on the drift 
velocity 
(typically 0.15 to 0.21~cm/\textmu s), the current induced on the anodes lasted 1.5 to 2~\textmu s. 

The waveforms measured on the grid and on the two anodes provided information about the location of
charge creation and Compton scattering if multiple charge pulses were observed. 
The time of charge arrival relative to the light signal gave information about the position along 
the drift direction of the electrons. 

\begin{figure}[!h]
\begin{center}
\includegraphics[width=\linewidth]{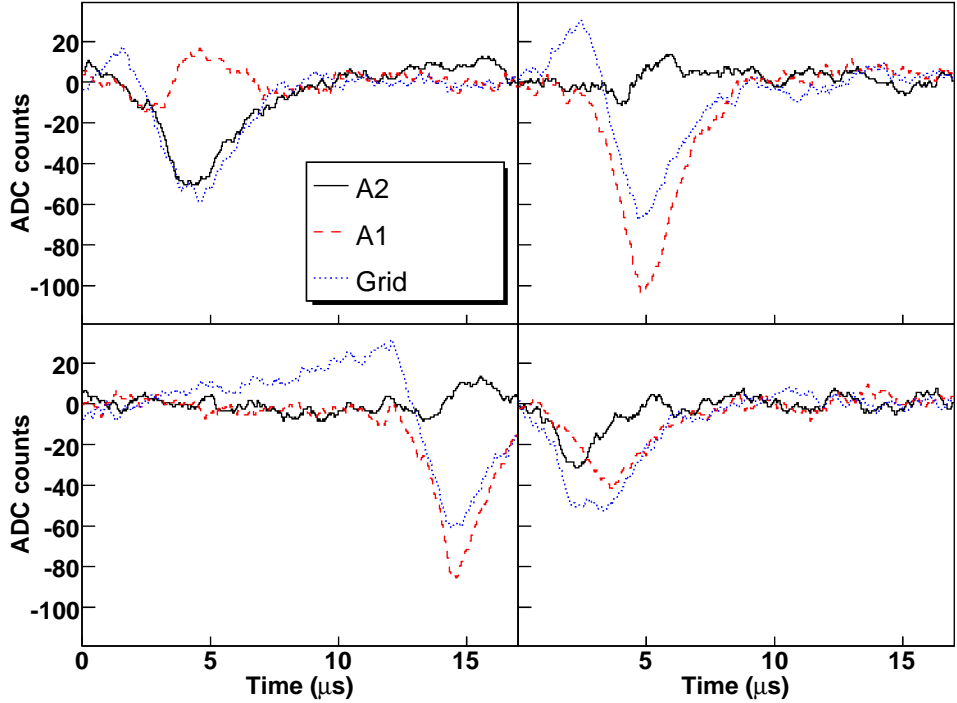}
\caption{Example of waveforms with a 2 kV/cm drift field. 
Central anode A1 (dashed line), peripheral anode A2 (solid line), grid (dotted line).}
\label{fig_wave}
\end{center}
\end{figure}

Figure~\ref{fig_wave} shows examples of four charge waveform events recorded by the 
20~MHz waveform digitizers chosen to illustrate several  types of events. 
In the upper two plots the charge was created roughly at the same z position; in the left plot the 
signal of the central anode (A1) integrated to zero and the anode A2 collected the charge, whereas in the 
right hand plot the interaction deposited the full charge on the central anode A1.
The lower left panel shows an event originating close to the cathode plane, resulting in a measured 
drift time of 15~\textmu s. The charge was collected by the central anode. The bipolar shape of the grid signal 
is clearly visible. The waveform measured on the grid depends on the z position of the interaction
and is also influenced by electrons collected by the grid. 
The lower right panel shows two photon interactions, presumably one Compton scattering 
and one photo-electric interaction. One interaction took place above A1 and one above A2. The interactions 
were also separated in the drift direction so that a two peak structure is visible in the
grid waveform. 
Simulations of the setup showed that only a small fraction (less than 5\%) of events fully contained on A1 have
mutliple hits that can be detected. The total charge for these events however is not significantly 
different from the events with just one interaction.
For this analysis we did not treat them separately since we were primarily interested in the total charge deposited on the anode.
Better separation of multiple photon interactions on an event by event basis will be 
possible with finer segmentation of the readout electrodes and shorter shaping time.

The purity of the LXe has an impact on charge collection.
In the current setup we achieved an electron lifetime of 200~\textmu s using purification in the gas 
phase with heated getters\footnote{A problem occurred with the 
purification system during the data taking with the NaI coincidence trigger used in this paper
resulting in  an electron lifetime of 90~\textmu s for much of the data presented here.}. 
We estimated that the level of purity obtained would result in a loss of 8\% of the electrons due to attachment.

For the analysis in this paper we selected events where no net charge was measured on A2.
By demanding the absence of charge on A2 the region of A1 in which events were 
accepted was smaller than its physical size since charge depositions close to the edge
of  A1 induced charge on A2.
The effective radius of the tube in which events were accepted was estimated to be
0.45~cm compared to the A1 radius of 0.5~cm.

\begin{figure}[!h]
\begin{center}
\includegraphics[width=0.7\linewidth]{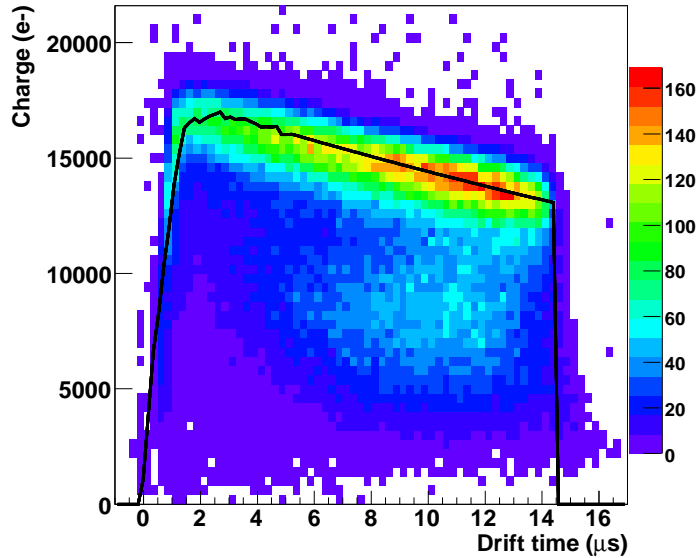}
\caption{Charge collection as a function of drift time for a 1~kV/cm drift field. 
The curve is a fit based on parametrization obtained from current calculations for energy deposits 
of 511 keV. The scale on the right corresponds to the number of events that occurred at a specific 
time with a certain charge deposition.}
\label{fig_charge}
\end{center}
\end{figure}

Figure~\ref{fig_charge} shows the distribution of charge due to 511 keV photons incident on the chamber 
as measured on A1 as a 
function of the drift time for a 1~kV/cm drift field. The shape of the distribution is the same for 
all drift fields. The 511~keV band rises sharply in less than 1~\textmu s, and then falls slowly until 
the cutoff which corresponds to the edge of the chamber. The sharp rise corresponds to photons 
interacting between the grid and anode. In that case the electronics, which is not sensitive to the 
charge induced by the much slower drifting ions,  measures only a fraction of the charge which is approximately 
proportional to the distance between the anode and the interaction point. When the interaction point is 
between the grid and the cathode, the measured charge 
should be independent of the interaction position. The decline of measured charge with increasing 
drift  times is due to electron attachment by impurities in the LXe.

Compton scattering interactions are evident below the 
511~keV band. They are due to photons entering the chamber with less than 511~keV  because they have 
scattered in the passive detector material, mostly the 2 cm of LXe between the vessel wall and the cathode, 
 and to photons escaping after a Compton scattering  
interaction in the liquid. 

We performed a fit of the 511~keV band to extract the drift velocity $v_d$, the total charge $Q_{tot}$ produced in the 
photon interaction and the attenuation length. 
The values obtained for these quantities are listed in Table~\ref{tab_Qresults} with their statistical 
uncertainties from the fits.
The charge yield $Q_{tot}/Q_0$ is also shown along with  $Q_{tot}$ which is the 
measured charge corrected for attachment and electronics calibration
and $Q_0$ is the ratio of the energy deposited by the $\gamma$-ray and the average 
energy to produce an electron ion pair: $Q_0=E_{\gamma}/W$ with $W$=15.6 eV~\cite{DokeTLXe}.

\begin{table}[!h]
\centering
\begin{tabular}{|c|c|c|c|c|}
\hline
$E_d$ & $v_d$  & $Q_{tot}$  & $\tau$ & \underline{$Q_{tot}$} \\
$\left[\mathrm{kV/cm}\right]$ & [cm/\textmu s] & (511~keV e$^-$) & [\textmu s] & $Q_0$\\\hline
0.33 & 0.16 $\pm$ 0.01 & 19 707 $\pm$  55   & 94 $\pm$ 3 & 0.60 \\
1    & 0.18 $\pm$ 0.01 & 23 372 $\pm$  59   & 61 $\pm$ 2 & 0.71 \\
2    & 0.20 $\pm$ 0.01 & 25 092 $\pm$  100  & 76 $\pm$ 5 & 0.77 \\
2.66 & 0.20 $\pm$ 0.01 & 24 761 $\pm$  35   & 60 $\pm$ 1 & 0.76 \\
\hline			 
\end{tabular}
\caption{Drift velocity ($v_d$),number of electrons 
($Q_{tot}$), electron lifetime $\tau$ and charge yield observed for different electric fields 
($E_d$).\label{tab_Qresults}}
\end{table}

\begin{figure}[!h]
\begin{center}
\includegraphics[angle=-90,width=0.9\linewidth]{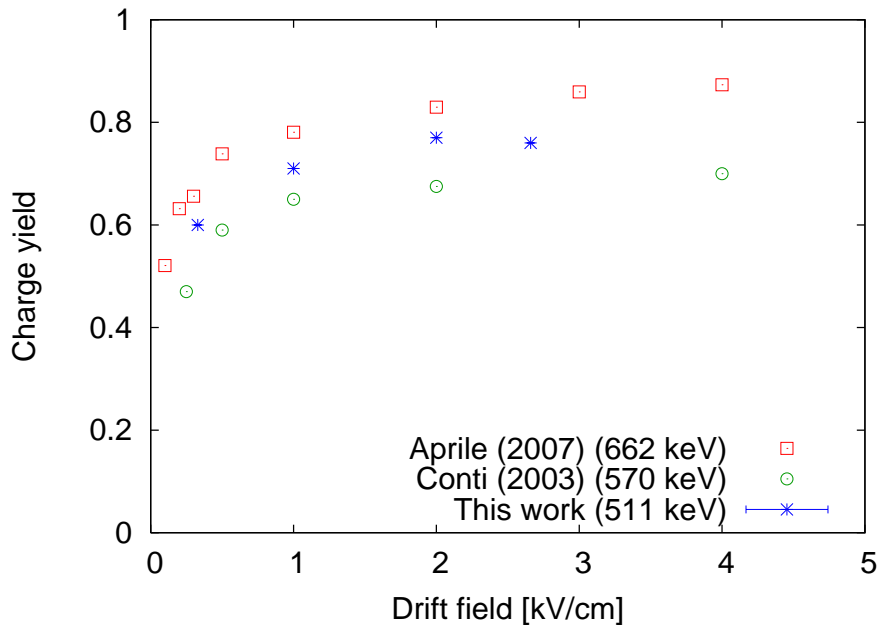}
\caption{Charge yield measured by different groups at different $\gamma$-ray energies: 
  this work marked with $\star$, \cite{columbia} with $\Box$ and \cite{stanford} with $\bigcirc$.}
\label{fig_Qyield}
\end{center}
\end{figure}

Figure \ref{fig_Qyield} shows the comparison of our results for the charge yield to the values obtained 
in \cite{columbia} and \cite{stanford}. Our results lie in between the two previous measurements 
of the charge yield. The obtained drift velocity was in agreement with previous measurements in \cite{vdrift1}.

\section{Light Collection}\label{sec_L}
\begin{figure}[!h]
\begin{center}
\includegraphics[width=0.7\linewidth]{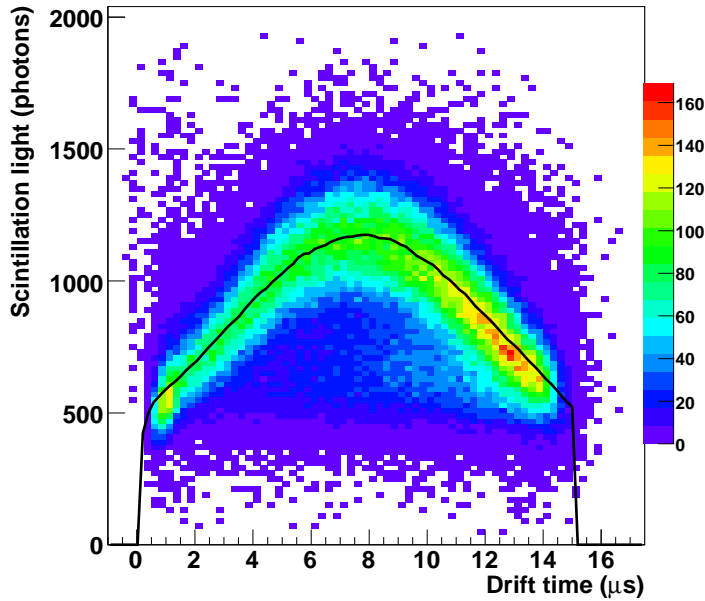}
\caption{Light collection as a function of drift time for a 1~kV/cm drift field at 511~keV. 
The curve is a fit based on parametrization obtained from solid angle calculations.
The scale on the right corresponds to the number of events that occurred at a specific 
time with a certain charge deposition.}
\label{fig_light}
\end{center}
\end{figure}

Scintillation light was detected by the LAAPDs located on two sides of the chamber. Figure~\ref{fig_light} 
shows the sum of the number of photons measured by both LAAPDs as a function of the
electron drift time for the events where all the charge was collected on the central anode.

The bell shape in Fig.~\ref{fig_light} is due to variations of the solid angle with drift distance  
which can be calculated by integrating over the 
LAAPD area for a given location in the chamber assuming no reflections occurred in the chamber walls.
The solid angle 
varied significantly with the position of the photon interaction. The arrival time of the electrons 
provided a handle on the solid angle variation in the drift direction. However, there was no information 
about the position of the interaction within the disk defined by A1. When the LAAPDs were 
used independently, the solid angle variation within this disk introduced a 22\% fluctuation in the light 
collection. Combining both LAAPDs reduced the fluctuation to 6\%.  

We fitted the distribution in fig.~\ref{fig_light} using a parametrization of the solid angle, with  
the total number of photons and the drift velocity as free parameters.  
The fit parameters are shown in Table~\ref{tab_Lresults}. 
The drift velocity is consistent with the one extracted from the fit to the charge distribution.  
The total number of photons drops with increasing drift 
voltage in agreement with previous measurements~\cite{columbia,stanford}. 
The number of photons actually created within the detector was not extracted in 
this analysis because we did not measure the photon detection efficiency (PDE) of the LAAPDs. 
One of the LAAPDs 
detected more photons than the other one, which suggests that there may be a variation of 
the PDE between LAAPDs\footnote{This may explain the apparent discrepancy between measurements made 
by different groups~\cite{LAAPD2}}. In the results given in Table~\ref{tab_Lresults}, 
we have assumed 100\% PDE for the 
LAAPD that exhibited a more stable operation and scaled the light measured by the other LAAPD 
accordingly introducing a systematic uncertainty
because of the unknown efficiency (which may be up to 50\%).
Uncertainties also originated from the fact that the ratio between the mean value measured by the two 
LAAPDs varied between data sets by 10\%. The light yield was computed using 
a value of 13.8~eV~\cite{DokeTLXe} needed to create one photon at zero drift field resulting 
in $S_0=37029$ photons.
$N_{tot}$ is the number of measured photons corrected for the solid angle of the geometry but not corrected 
for the photo detection efficiency of the LAAPDs.

\begin{table}[!h]
\centering
\begin{tabular}{|c|c|c|c|}
\hline
$E_d$ [kV/cm] & $v_d$ [cm/\textmu s] & $N_{tot}$ (511~keV e$^-$) & $N_{tot}/S_0$ \\
\hline
0.33 & 0.15 $\pm$ 0.01 & 12 161 $\pm$ 1269 & 0.33 \\
1    & 0.18 $\pm$ 0.01 & 10 113 $\pm$ 1055 & 0.27 \\
2    & 0.20 $\pm$ 0.01 & 9243  $\pm$ 964  & 0.25 \\
2.66 & 0.21 $\pm$ 0.01 & 7936  $\pm$ 828  & 0.21 \\
\hline
\end{tabular}
\caption{Electric field ($E_d$), drift velocity ($v_d$), number of photons 
($N_{tot}$) and light yield observed (see text) for 511~keV photon interactions.\label{tab_Lresults}}
\end{table}

\begin{figure}[!h]
\begin{center}
\includegraphics[angle=-90,width=0.9\linewidth]{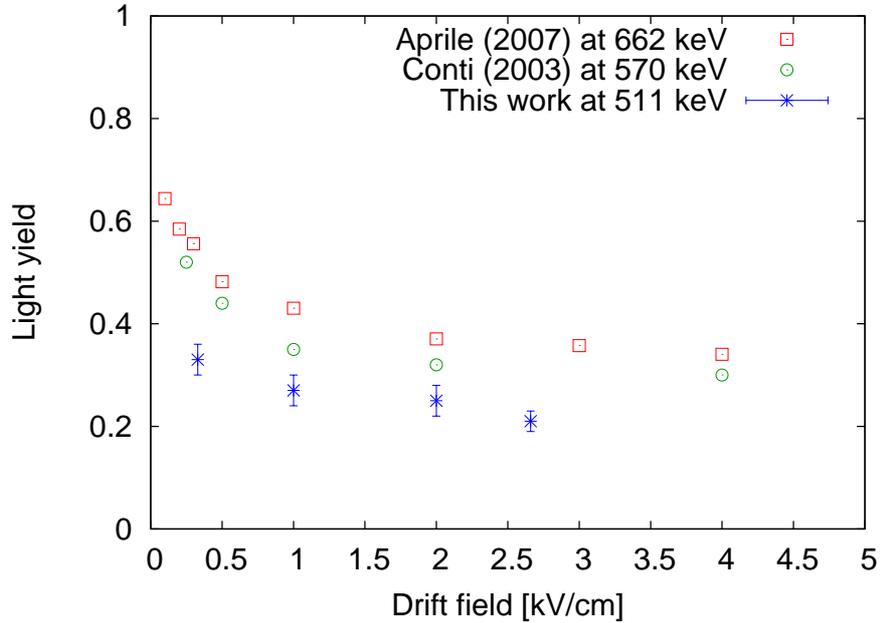}
\caption{Light yield relative to the maximum yield measured by different groups at different $\gamma$-ray energies: 
 this work marked with $\star$, \cite{columbia} with $\Box$ and \cite{stanford} with $\bigcirc$.}
\label{fig_Lyield}
\end{center}
\end{figure}

Figure~\ref{fig_Lyield} shows the comparison of our results with values obtained in 
\cite{columbia} and \cite{stanford}. If the quantum efficiency of the LAAPDs was 60\%, which later results
presented here suggest, our results would be in agreement with previous measurements.

\section{Light and Charge Combination}\label{sec_E}
\begin{figure}[!h]
\begin{center}
\includegraphics[width=\linewidth]{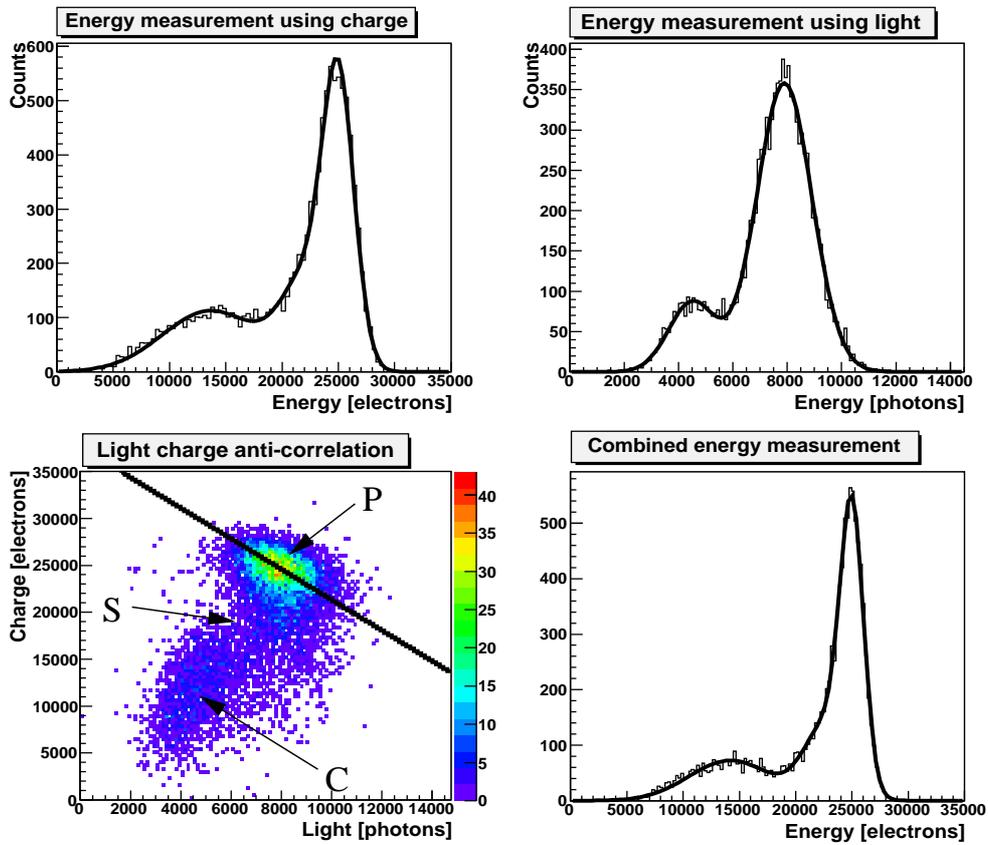}
\caption{The observed charge spectrum (upper left plot), light spectrum (upper right plot), 
correlation between light and charge signals (lower left plot), and combined spectrum using 
the correlation (lower right plot) for 511~keV photons with a drift field of 2.66~kV/cm. 
The data points in the correlation plot (lower left) that are not part of the Compton (C) or the 
photoelectric peak 
(P) are due to photons that scattered outside the detector (S). The linear fit (solid line) giving the axis 
of the correlation ellipse is depicted as well. The fits shown (solid lines) were made with a sum of 3 Gaussians 
(upper left and lower right plot) or 2 Gaussians (upper right plot).}
\label{fig_energy}
\end{center}
\end{figure}

To study the energy resolution we focused on the central region of the chamber by 
selecting events with no charge on A2 and choosing a time window in the drift direction corresponding to 
2~mm drift located on the axis of the  LAAPDs where the light collection is maximal as shown in 
Fig.~\ref{fig_light}. The charge signals were corrected for attenuation and the light signals for the 
difference between the two LAAPDs and the solid angle dependence in the drift direction. 
Resolution results are given as the standard deviation ($\sigma$) of a Gaussian distribution unless 
otherwise stated.
Figure~\ref{fig_energy} shows the analysis of a data run at a drift field of 2.66~kV/cm.
Evaluating the charge and light signals separately gave energy resolutions of 12.1\% for light and 
5.4\% for charge by fitting the spectra (shown in the upper plots of Fig.~\ref{fig_energy}) with a sum of two 
Gaussians and evaluating the mean and width of the 511 keV peak. 
The energy resolution can be 
improved significantly by combining the information from light and charge using the anti-correlation 
of the two signals~\cite{columbia,stanford,Doke}. 
The lower left plot of Fig.~\ref{fig_energy} shows the linear anti-correlation between the light and 
charge measurement and the axis of the ellipse. 
Selecting the 511~keV region of the photo-electric-peak the correlation angle was 
obtained from a linear fit which provided the axis of the charge-light ellipse. 
Projecting the data points along this axis as described in~\cite{columbia} gave the overall energy 
resolution. The correlation angle given here depended on the detector geometry and the 
efficiency to measure light and charge separately. 
The upper left plot shows the charge spectrum collected on the anode which is equal to a projection of the 
correlation along the light axis. In the upper right plot the projection of the correlation along the 
charge axis can be seen, giving the spectrum of the collected light.  
The lower right plot demonstrates the improved energy resolution of the combined spectrum when 
projecting along the correlation axis and normalizing to the mean charge. The sum of three Gaussians 
was used as the fit function to account for the three contributions
to the spectrum: the Compton region (C), the photoelectric peak (P), and scattered events (S) which lost 
energy outside the detector, mostly in the LXe before entering the chamber.

Another variable to quantify the anti-correlation between light 
and charge is the correlation coefficient $\rho$~\cite{rho}.  
Assuming that the probability for a recombining
electron-ion pair to produce a scintillation photon is 1 and the detector would be able 
to measure light and charge with 100\% efficiency and perfect resolution, 
$\rho$ should be -1. Deviation from -1 could be due to other sources of fluctuations
like density fluctuations or delta electrons as discussed in~\cite{fluctuations} and the references within.

Table~\ref{tab_Eresults} gives the results of the analysis for different drift fields. 
The best combined energy 
resolution reached for these data sets was 4.1\% at 2.66~kV/cm drift field (see below).

\begin{table}[!h]
\centering
\begin{tabular}{|c|c|c|c|c|c|}
\hline
$E_d$ & \multicolumn{3}{c|}{Energy resolution [\%] } & $\theta_{corr}$ & $\rho$\\
$\left[\mathrm{kV/cm}\right]$ & light &  charge & combined & [$^\circ$] & \\ \hline
0.33 &	13.5  $\pm$ 0.2 & 7.3 $\pm$ 0.5	& 4.7 $\pm$ 0.1	 & 56 &	-0.46\\
1    &	12.2  $\pm$ 0.2 & 6.0 $\pm$ 0.3	& 4.3 $\pm$ 0.3  & 59 &	-0.34\\
2    &	12.8  $\pm$ 0.5 & 7.0 $\pm$ 0.6	& 4.8 $\pm$ 0.4  & 62 &	-0.34\\
2.66 &	12.1  $\pm$ 0.1 & 5.4 $\pm$ 0.2	& 4.1 $\pm$ 0.1  & 58 &	-0.26\\
\hline
\end{tabular}
\caption{Energy resolutions ($\sigma$) observed at different drift fields for light and charge 
separately and combined result using the correlation.\label{tab_Eresults}}
\end{table}

\section{Discussion of Error Sources}\label{errsection}
In this section we discuss contributions to the energy resolution that were due to detector inefficiencies 
or physics constraints like light-charge-fluctuations. When a photon interacts there is an 
initially produced number of ionization charges and scintillation photons which is modified by 
recombination dependent on the presence of an applied electric field.
Table~\ref{tab_Evariables} summarizes the variables used in the calculation of error contributions.

\begin{table}[!h]
\centering
\begin{tabular}{|c|c|c|}
\hline
& Charge & Scintillation light \\ \hline
Initial & $Q_i$     & $S_i$ \\
Final  & $Q_f$ = $Q_i [1-F_r(E_d)]$ & $S_f = S_i + Q_i F_r(E_d) P_{e \rightarrow h\nu}$\\
Measured & $Q_m$ = $A \ Q_f$ & $S_m = F_{\Omega} \epsilon S_f$\\
\hline
\end{tabular}
\caption{Parameters used in the discussion of energy resolution as described in the text.\label{tab_Evariables}}
\end{table}

$Q_i, Q_f$ and $Q_m$ ($S_i, S_f$ and $S_m$ ) are the numbers of initially produced, post-recombination, 
and  measured charge 
(light) signals respectively. $F_r(E_d)$ is the fraction of electron-ion pairs that recombine for a given electric 
field $E_d$, and 
$P_{e \rightarrow h\nu}=1$ \cite{columbia} gives the probability for a recombining electron-ion pair 
to produce a 
scintillation photon. Impurities may capture some electrons, which is accounted for by an attenuation 
parameter $A$, 
which depends on the electron drift distance.
The photo-detectors have a  photo-detection efficiency $\epsilon$ 
and cover a fraction of the total solid angle $F_{\Omega}$.

\subsection{Charge}
The charge resolution is dominated by electronics noise and charge-light fluctuation. 
The charge-light fluctuation is expressed by the fluctuation of the recombination fraction $F_r$, $\Delta F_r$. 
The charge resolution can then be written as

\begin{equation}
\left(\frac{\Delta Q_m}{Q_m}\right)^2 = 
\left(\frac{ENC_q}{Q_m}\right)^{2} +
\left(\frac{\Delta F_r}{1-F_r}\right)^2 + \frac{1-A}{Q_m}
\end{equation}

where the first term on the right describes the electronics noise of the amplifier, the second term quantifies the 
light-charge fluctuation, and the third contribution is the attachment factor which is negligible.
$\Delta F_r$ describes the fluctuation of the recombination and will also occur in the discussion of the
error sources for the light measurement in the next section.
Table~\ref{tab_QEsources} gives values for the error contributions to the energy 
resolution obtained from the 
charge measurement. Also shown is the intrinsic energy resolution found by subtracting  the noise of the 
electronics from the measured
resolution. The only unknown variable contributing to the 
error of the energy resolution is $\Delta F_r$ which can be calculated once the intrinsic energy 
resolution
is known. The values for $\Delta F_r$ obtained can also be found in Table~\ref{tab_QEsources}.

\begin{table}[!h]
\centering
\begin{tabular}{|c|c|c|c|c|}
\hline
$E_d$ & Measured & Noise &  Intrinsic & $\Delta F_r$\\
$\left[\mathrm{kV/cm}\right]$ & res. [\%]& [\%] & res. [\%] & [\%]\\\hline
0.33 & 7.31 $\pm$ 0.54 & 5.03 $\pm$ 0.04 & 5.3 $\pm$ 1.5 & 3.2 $\pm$ 0.9\\
1    & 6.04 $\pm$ 0.33 & 4.22 $\pm$ 0.03 & 4.3 $\pm$ 0.9 & 3.1 $\pm$ 0.7\\
2    & 7.00 $\pm$ 0.62 & 4.29 $\pm$ 0.03 & 5.5 $\pm$ 1.6 & 4.2 $\pm$ 1.2\\
2.66 & 5.43 $\pm$ 0.17 & 3.48 $\pm$ 0.03 & 4.2 $\pm$ 0.4 & 3.2 $\pm$ 0.3\\
\hline
\end{tabular}
\caption{Contribution of error sources to the energy resolution obtained from the charge measurement.\label{tab_QEsources}}
\end{table}

\subsection{Light}
The fluctuations in the LAAPD specified as the excess noise factor $F(M)=2+kM$ is dependent on the gain 
$M$ and affects the resolution. For this setup with $k=0.001$, $F(M)=2.5$.
Furthermore, the LAAPD gain $<10^3$ requires that low noise electronics must be used to further amplify 
the signal, which adds electronic noise $ENC_s$. Another source of fluctuations arises 
because the solid angle seen by the photo-sensor may vary on an event-by-event basis since 
the solid angle changes with the position of the photon interaction within A1. This fluctuation can be 
corrected for if the interaction position is known well from the ionization signal.

Neglecting other detection fluctuations, the light signal resolution for our setup can be written as:

\begin{eqnarray}
\left(\frac{\Delta S_m}{S_m}\right)^2 &=& \left(\frac{ENC_s}{M S_m}\right)^2 + \frac{F(M)}{S_m} \nonumber \\ 
&+&\left(\frac{\Delta F_{\Omega}}{F_{\Omega}}\right)^2 +
\left(\frac{P_{e \rightarrow h\nu}Q_i \Delta F_r}{S_f}\right)^2\\
&+&\frac{F_r Q_iP_{e \rightarrow h\nu}(1-P_{e \rightarrow h\nu}) }{S_f^2} \nonumber
\end{eqnarray}

where the first term on the right represents the electronics noise, the second term  gives the 
contribution from fluctuations in the LAAPD gain,
the third term is the fluctuation of the solid angle due to the position of the light creation inside 
the chamber and the fourth 
term describes the light-charge fluctuation. The contribution of the fluctuation in  
$P_{e \rightarrow h\nu}$ given by the last term is negligible or exactly zero if $P_{e \rightarrow h\nu}=1$.
Table~\ref{tab_LEsources} gives values for the error contributions to the measured energy
resolution from the 
scintillation light. The solid angle fluctuation amounted to 5.6\% and was independent of the drift 
field.
Also shown is the intrinsic energy resolution found when subtracting those error sources due to the 
detector from the measured resolution. 
$\Delta F_r$ was calculated again and can be compared to the values obtained from the charge measurement. 
The values for $\Delta F_r$ from both the light and charge measurements are in good agreement  within statistical errors
providing a consistency check for the error analysis.

\begin{table}[!h]
\centering
\begin{tabular}{|c|c|c|c|c|c|}
\hline
$E_d$ & Measured  & Noise & LAAPD & Intrinsic & $\Delta F_r$\\
$\left[\mathrm{kV/cm}\right]$ & res. [\%] & [\%] & fluct. [\%] &  res. [\%] & [\%]\\ \hline
0.33 & 13.5 $\pm$ 0.2 & 3.3  & 0.46 &  9.6 $\pm$ 0.4 & 5.8 $\pm$ 1.3\\
1    & 12.2 $\pm$ 0.2 & 4.0  & 0.55 &  6.8 $\pm$ 0.8 & 3.3 $\pm$ 1.0 \\
2    & 12.8 $\pm$ 0.5 & 5.1  & 0.56 &  7.1 $\pm$ 1.7 & 3.1 $\pm$ 1.2 \\
2.6  & 12.1 $\pm$ 0.1 & 4.7  & 0.63 &  5.5 $\pm$ 0.5 & 2.5 $\pm$ 0.8 \\ 
\hline
\end{tabular}
\caption{Contribution of error sources to the energy resolution obtained using
  the light measurement.\label{tab_LEsources}}
\end{table}

\subsection{Combination}
Combining the light and charge allows improvement to the resolution by canceling the fluctuations 
of $F_r$ 
by making use of the anti-correlation, provided the measured charge is corrected for attenuation and 
the measured light for solid angle and PDE: 

\setcounter{equation}{0}
\renewcommand{\theequation}{3\alph{equation}}
\begin{eqnarray}
E_c&=&\frac{Q_m}{A}+\frac{S_m}{F_{\Omega}\epsilon}\label{eq1}  \\
&=&Q_f+\frac{S_f}{P_{e \rightarrow h\nu}} \label{eq2}\\
&=&Q_i\left(1-F_r\right)+\frac{S_i}{P_{e \rightarrow h\nu}}+Q_i F_r \label{eq3}\\
&=&Q_i+\frac{S_i}{P_{e \rightarrow h\nu}} \label{eq4}
\end{eqnarray}
\renewcommand{\theequation}{\arabic{equation}}
\setcounter{equation}{3}

where $E_c$ is the energy measured by combining the charge and light signals:
In eq.~\ref{eq1} the light and charge signals were combined; eq.~\ref{eq2} and eq.~\ref{eq3}
made use of the formulas in Table~\ref{tab_Evariables}; and in eq.~\ref{eq4} $F_r$ was eliminated.
The remaining uncertainty in the combined energy resolution is then:

\begin{eqnarray}
\Delta E_c^2 &=& 
\frac{1}{\epsilon^2 F_{\Omega}^2}\left[\left(\frac{ENC_s}{M}\right)^2 +
F(M) S_m + 
\left(\frac{\Delta F_{\Omega} S_m}{F_{\Omega}}\right)^2\right] \nonumber\\
&+& \left(\frac{ENC_q}{A} \right)^2 \\ 
&+&\frac{Q_f(1-A)}{A} + \frac{F_rQ_i(1-P_{e \rightarrow h\nu})}{P_{e \rightarrow h\nu}}\nonumber
\end{eqnarray}

where the first term on the right hand side (in brackets) originated from the light signal contribution and the 
second term from the charge measurement.
The last two terms describing the contributions from the binomial statistics of $A$ and 
$P_{e \rightarrow h\nu}$ are negligible.

Table~\ref{tab_CEsources} gives the calculated values for the combined energy resolution $\Delta E_c/E_c$. 
The intrinsic combined energy resolution given in the last column of table~\ref{tab_CEsources} was 
obtained by subtracting $\Delta E_c/E_c$in quadrature from the measured values. The solid angle was
12\% and the efficiency $\epsilon$ for the LAAPDs was assumed to be 1. 

\begin{table}[!h]
\centering
\begin{tabular}{|c|c|c|c|}
\hline
$E_d$ & Meas. comb. & $\Delta E_c/E_c$ & Intr. comb. \\
$\left[\mathrm{kV/cm}\right]$ & res [\%] & [\%] & res. [\%] \\\hline 
0.33 & 4.74 $\pm$ 0.09 & 4.0 $\pm$ 0.1 &  2.5 $\pm$ 0.4 \\
1    & 4.31 $\pm$ 0.26 & 3.7 $\pm$ 0.1 &  2.3 $\pm$ 1.0 \\
2    & 4.78 $\pm$ 0.35 & 3.8 $\pm$ 0.1 &  2.9 $\pm$ 1.2 \\
2.66 & 4.14 $\pm$ 0.10 & 3.3 $\pm$ 0.1 &  2.5 $\pm$ 0.4 \\
\hline
\end{tabular}
\caption{Contribution of error sources and corrected intrinsic energy resolution for combined 
charge-light measurement.\label{tab_CEsources}}
\end{table}

The intrinsic correlation coefficient was calculated but showed a large
uncertainty due to the impact of the uncertainty on $\Delta F_r$.

Another consistency check of the formulas presented here can be made by extracting the factor
$\epsilon P_{e \rightarrow h\nu}$ in two different ways from the data. We cannot disentangle the 
efficiency from
the probability but the product can be obtained from the correlation angle of the 511 keV 
cloud by writing $Q_m/A$ as a function of $S_m/F_{\Omega}$ with the slope 
$m$ describing the axis of the ellipse:

\begin{eqnarray}
\frac{Q_m}{A}=Q_f=m \frac{S_m}{F_{\Omega}}+const.\\
\Longrightarrow \theta_{corr}=\arctan{\left(\frac{-1}{ \epsilon P_{e \rightarrow h\nu}}\right)} \nonumber
\end{eqnarray}

The results are listed in Table~\ref{tab_Pepsilon}. The mean value over all runs was
$\epsilon P_{e \rightarrow h\nu}=0.60 \pm 0.03$.

\begin{table}[!h]
\centering
\begin{tabular}{|c|c|c|}
\hline
$E_d \left[\mathrm{kV/cm}\right]$ & $\theta_{corr}$ & $\epsilon P_{e \rightarrow h\nu}$\\\hline
0.33 & 56.2 & 0.67 $\pm$ 0.03\\
1    & 58.7 & 0.61 $\pm$ 0.11\\
2    & 62.3 & 0.53 $\pm$ 0.30 \\
2.66 & 58.5 & 0.61 $\pm$ 0.05\\ 
\hline
\end{tabular}
\caption{Calculating $\epsilon P_{e \rightarrow h\nu}$ on event by event basis.\label{tab_Pepsilon}}
\end{table}

\begin{figure}[!h]
    \begin{center}
      \includegraphics[width=0.95\linewidth]{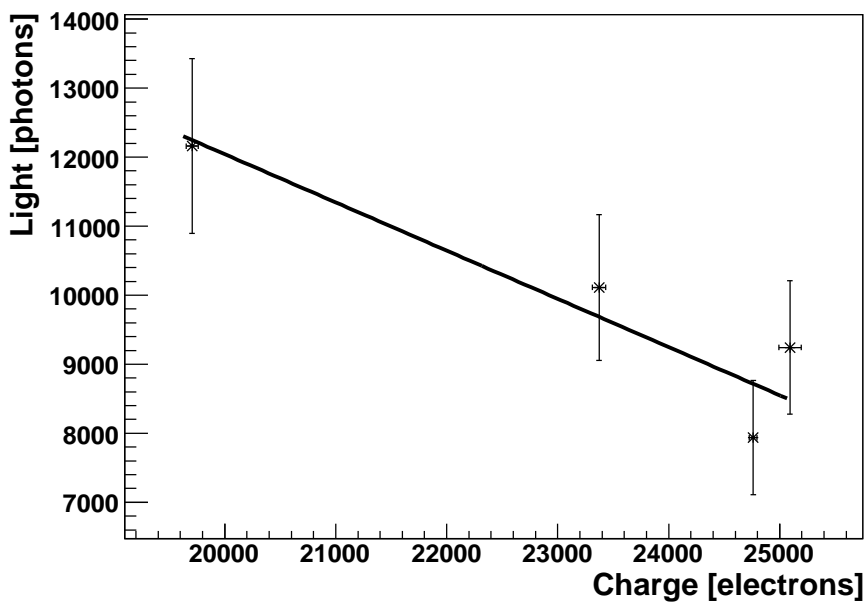}
      \caption{Method of extracting $\epsilon P_{e \rightarrow h\nu}$ from mean values of the 
	correlation cloud parameters described in the text. The solid line is the linear fit to the data.\label{fig_Pe}}
    \end{center}
   \end{figure}

The other method for extracting $\epsilon P_{e \rightarrow h\nu}$ is to find the slope from the plot of the mean 
values of light vs. charge (coordinates  of the center of the ellipse) for 
each electric field setting.
This is shown in Fig.~\ref{fig_Pe}. The linear fit to this data  gave a slope of 
$\epsilon P_{e \rightarrow h\nu}=0.7 \pm 0.3$ 
which  agrees with the value obtained from the event-by-event method.
This result would be consistent with $P_{e \rightarrow h\nu}=1$ if the efficiency of the APDs 
were 60\%.

\subsection{Improving Light Resolution using Position Reconstruction}
In the current system, the energy resolution contribution from light was partially limited by the
 uncertainty in the position of the light source due to the 1 cm dia. size of A1 i.e.  from  the fluctuation 
of the solid angle within A1.  Applying the formulas presented above
to a Geant4~\cite{geant4} simulation, it was found  that knowing the position of the light signal to 1 mm 
would improve the light resolution by about 1.5\% and the combined resolution by up to 0.5\% 
giving 10\% for light alone and 3.6\% for the combination of charge and light,
consistent with the resolution reported in~\cite{columbia}.
Table~\ref{tab_simresults} summarizes the values obtained with the simulation comparing the two 
cases of not knowing the position of the interaction within A1  and being able to correct for it.

\begin{table}[!h]
\centering
\begin{tabular}{|c|c|c|c|}
\hline
 & Light res. [\%] &  Charge res. [\%] & Combined res. [\%] \\ \hline
Measured  &	12.1  $\pm$ 0.1 & 5.4 $\pm$ 0.2	& 4.1 $\pm$ 0.1 \\
Simulated &     12.0  $\pm$ 0.2 & 5.4 $\pm$ 0.1	& 3.9 $\pm$ 0.1 \\
Corrected &     10.4  $\pm$ 0.2 & 5.4 $\pm$ 0.1& 3.6 $\pm$ 0.1  \\
\hline
\end{tabular}
\caption{Energy resolutions obtained from the simulation with (corrected) and without (simulated) the 
  correction for the position of the interaction within A1 in comparison with the measured resolutions.
  The charge resolution was not affected by the correction.\label{tab_simresults}}
\end{table}

\section{Summary and Conclusion}
\begin{figure}[!h]
    \begin{center}
      \includegraphics[height=0.8\linewidth,angle=-90]{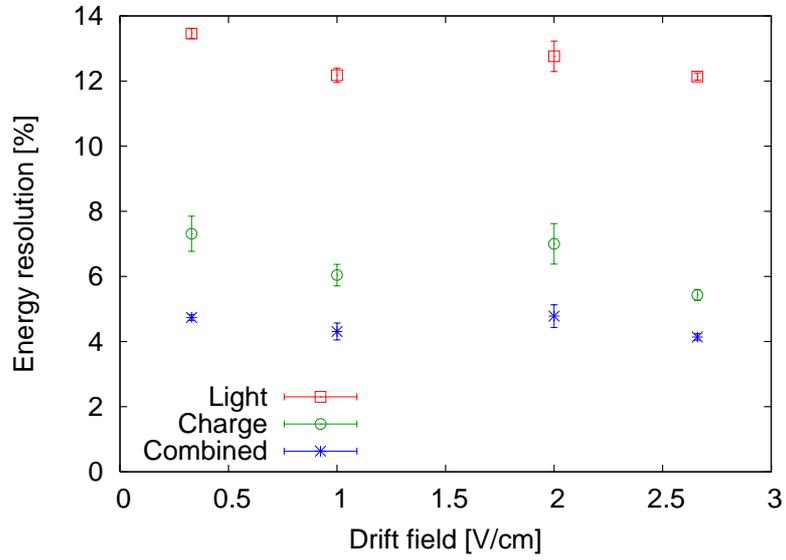}
      \caption{Energy resolution from charge ($\bigcirc$) and light ($\Box$) measurements as well as the combined 
	($\star$) resolution  for different drift fields. \label{fig_Eres}}
    \end{center}
\end{figure}

Measurements have been made of the response of a liquid xenon drift chamber to irradiation by 511~keV 
photons. Using a model accounting for the sources of uncertainty in the energy 
resolution we also determined values for the intrinsic energy resolutions. 
Figure~\ref{fig_Eres} summarizes the results for charge, light and combined energy resolution as a function
of the drift field. The main error contribution to the combined energy resolution, apart from the solid 
angle fluctuations which can be eliminated by utilizing the position measurement, originated from the APD 
gain fluctuation and the anode noise. Both were about of 2.7\%.

\begin{figure}[!h]
  \begin{center}
    \includegraphics[height=0.8\linewidth,angle=-90]{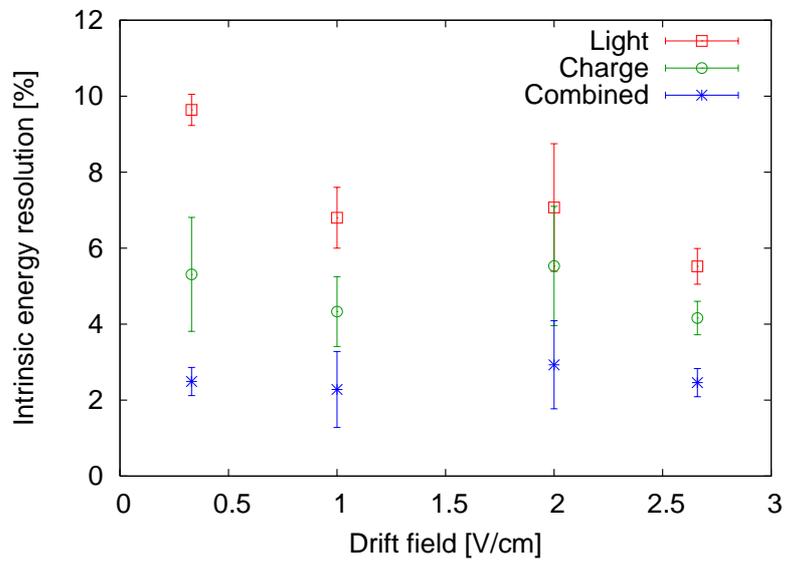}
    \caption{Intrinsic energy resolution from charge ($\bigcirc$) and light ($\Box$) measurements as well as the 
      combined ($\star$) intrinsic resolution for different drift fields. \label{fig_IntEres}}
  \end{center}
\end{figure}

Figure~\ref{fig_IntEres} depicts the values for the intrinsic energy 
resolutions obtained by subtracting the detector contributions from the values in fig.~\ref{fig_Eres}.
The error bars given are statistical.

Based on these results, the combined energy resolution of $<$3.5\% (or $<8\%$ FWHM) would be anticipated 
in a detector configuration suitable for applications to PET which would have comparable light collection
efficiency to the prototype detector described above and $<$1~mm spatial resolution.
Reducing the anode to grid spacing to 1~mm and 
the grid wire spacing to  1~mm  will reduce the width of the pulses and minimize the dependence of the 
pulse shape on the location of the electron cloud. 
Further improvements are foreseen in areas including purification and low noise electronics.

\section*{Acknowledgments}
We  thank R. Bula, M. Constable, and  C. Lim for their technical contributions to this work
and P. Gumplinger for assistance with simulations. We also thank E. Aprile and E. Conti
for providing information about their work. 
This work was supported in part by the Canada Foundation for Innovation, the University of British Columbia,  
and  TRIUMF which  receives federal funding via a contribution agreement through the National Research Council 
of Canada.

\end{document}